# Continuous Models of Epidemic Spreading in Heterogeneous Dynamically Changing Random Networks

S.V. Ivanov[1], A.V. Boukhanovsky[1] and P.M.A. Sloot[1,2,3]

[1] National Research University of Information Technologies,

Mechanics and Optics (ITMO), Kronverkskiy 49, 197101 Saint Petersburg, Russia

[2] Computational Science, University of Amsterdam,

Science Park 904, 1098 XH Amsterdam, The Netherlands

[3] Nanyang Technological University, 50 Nanyang Avenue, 639798 Singapore

**Abstract** Modeling spreading processes in complex random networks plays an essential role in understanding and prediction of many real phenomena like epidemics or rumor spreading. The dynamics of such systems may be represented algorithmically by Monte-Carlo simulations on graphs or by ordinary differential equations (ODEs). Despite many results in the area of network modeling the selection of the best computational representation of the model dynamics remains a challenge. While a closed form description is often straightforward to derive, it generally cannot be solved analytically; as a consequence the network dynamics requires a numerical solution of the ODEs or a direct Monte-Carlo simulation on the networks. Moreover, Monte-Carlo simulations and ODE solutions are not equivalent since ODEs produce a deterministic solution while Monte-Carlo simulations are stochastic by nature. Despite some recent advantages in Monte-Carlo simulations, particularly in the flexibility of implementation, the computational cost of an ODE solution is much lower and supports accurate and detailed output analysis such as uncertainty or sensitivity analyses, parameter identification etc. In this paper we propose a novel approach to model spreading processes in complex random heterogeneous networks using systems of nonlinear ordinary differential equations. We successfully apply this approach to predict the dynamics of HIV-AIDS spreading in sexual networks, and compare it to historical data.

## 0 Introduction

Spreading processes in complex networks are modeled such that each node in the network represents an agent that can be in one of a set of finite number of states. As time is discrete, the next state of the agent changes at each time step according to its own state and the states of its neighbors in the network. Currently this approach is widely used in modeling and understanding the dynamics of spreading phenomena in for instance epidemic and rumor networks (Newman 2002,2003; Xie et al. 2011; Meyers et al. 2005). Even though these processes are quite different they may be described by the same method. In these models the emphasis is on the role of the network topology in determining the rate of the

e-mail: svivanov@mail.ifmo.ru; avb_mail@mail.ru; p.m.a.sloot@uva.nl
(Corresponding Author)

spreading process. Contact patterns are an important aspect of the heterogeneity within a collection of nodes; they refer to heterogeneous connectivity (i.e. the degree distribution is not uniform), individual nodes behavior, community structures, spatial networks properties and many others. Normally a given network is assumed to be static, i.e. edge dynamics is not considered. Such static networks may be a realistic model of population-wide average dynamics, it has the advantage of being well understood mathematically. For instance, in the case of an epidemic an exact solution for the mean outbreak time and expected final size of the epidemics is well understood (Meyers et al. 2005; Newman 2002; Volz 2008). But even for simple models the epidemic incidence dynamics, that is the number of newly infected individuals at time step t, is difficult to approximate without stochastic simulations. A new problem that was solved is a continuous approximation of infection rate of susceptible nodes having shared edges with different types and different numbers of infected nodes. This is an important issue for many real epidemiological processes when the probability to be infected strongly depends on number of contacts with infected individuals and probability of contamination during one contact is rather small (HIV for example).

The main factor that prevents successful modeling of realistic dynamical aspects of network epidemics and related processes is the heterogeneity of the underlying networks. Previous work used pair approximation or moment closure methods (Bauch 2002; Eames and Keeling 2002), the main disadvantage however being the high-dimensionality and extreme computational cost. Another aspect never considered in the exact solution of network dynamics is the changing of network links over time, an essential aspect of real-world networks (infection rate per contact rather than per nodes degree).

In this paper we present a new approach using a system of nonlinear ordinary differential equations that overcomes the problems described above and that allows for temporal reconfiguration of the network links. This model may be used to estimate the majority of process parameters that can be derived using numerical methods. We show the applicability of this method by calculating epidemic HIV-AIDS parameters and validating it against publicly available AIDS data for homosexual and heterosexual populations in the USA.

# 1 Spreading process in random network

## 1.1 Formal description of the model

Let us consider a Complex Network model (CN-model) as a set of the pairs $\{G, \Im\}$, where $G$ is a graph, that is, an ordered pair of disjoint sets $(V, E)$ (vortices and edges), and $\Im$ is an evolutionary operator, governing network changes in discrete time $t$:

$$\begin{aligned} \langle V, E \rangle_{t+1} &= \Im \langle V, E \rangle_t; \\ \langle V, E \rangle_{t=0} &\stackrel{def}{=} \langle V_0, E_0 \rangle. \end{aligned} \quad (1)$$

The evolutionary operator in Eq. 1 can be represented as a composition of distinctive operators $\Im = \bigotimes_k \Im_k$ corresponding to different dynamical aspects. The basic dynamical aspect is a state (virus, information etc) propagating over the network. At each time-step a given node can transfer its state to neighboring nodes with a given probability. This process may be represented by the percolation operator $\Im_1$. The simplest case of this model is a static network, i.e. the network structure doesn't change over time. A more elaborate model takes into account the network dynamics. This reflects the fact that network links are not stable and the network structure is evolving over time. The model is in principle fully connected, since in the case of even small number of links (for instance, only one for each node) the possibility to be connected to a node with given state cannot be disregarded. This process may be represented by the network dynamics operator $\Im_2$. Furthermore we define an operator $\Im_3$ that represents the node evolution, the way it changes its state over time.

## 1.2 A simple SIR model

Let's consider a network model of disease propagation over a network (Anderson and May 1991). The main idea of this model is to divide a population into three groups of susceptible, infective and removed individuals. The model allows for studying the dynamics of those groups as a result of contact between infected and healthy nodes through shared edges and natural processes of depletion of infected nodes. The basic algorithm can be written as follows:

1. Generate a network with given statistical properties and an initial number of randomly infected nodes $\rho_0$;

2. Infect nodes surrounded by infected nodes with probability $\lambda$ for every link;
3. For each infected node apply a rule of disease progression.
4. Remove dead or immunized nodes from the network;
5. Apply additional rules (treatment, demographic changes etc);
6. Store the current node states and rewire by generating a new random network (links dynamics);

Strictly spoken, stages 5-6 are not required but must to be taken into account for real-world processes. Every step of the algorithm can be implicitly interpreted as an independent operator $\Im_i$, introduced in section 1.1. For quantitative analysis of the results we need to repeat steps 2-6 using different initial conditions as long as the variance of simulation result is larger than a predefined threshold (i.e. reach statistical significance for the moments of the distribution). The resulting number of iterations is not known in advance and can be very large. It is possible to avoid this problem with the classical SIR models that are normally described in terms of a system of ordinary differential equations:

$$\frac{ds(t)}{dt} = -\lambda \rho(t) s(t)$$
$$\frac{d\rho(t)}{dt} = -\mu \rho(t) + \lambda \rho(t) s(t), \qquad (2)$$
$$\frac{dr(t)}{dt} = \mu \rho(t)$$

where $s(t)$ – *susceptible*, $\rho(t)$ – *infective*, $r(t)$ – *removed*. Susceptible individuals are capable of contracting the disease and becoming infected, infective individuals are capable of transmitting the disease; removed individuals have had the disease and are dead. Parameters $\lambda$ and $\mu$ are positive constants representing infection and removal rates, defined as the probability per unit time that an infective individual will pass the disease onto a particular susceptible network neighbor (Newman 2003). The value of $\lambda$ is a characteristic of the operator $\Im_1$ only and is determined by the details of the infection spreading. The value of $\mu$ depends on the disease evolution inside an infected node before its death or isolated status in which it is not infectious anymore. This model of course is too simple to be useful in real applications but it is a good starting point to develop more complicated and realistic models. Moreover this ODE model may be interpreted as a dynamical network model with a homogeneous structure (i.e. exactly one link per node), with fully reconfigured links at each time step. Taking into account that $\rho(t)$ and $s(t)$ are fractions of nodes, their product may be treated as the percentage of nodes in the network having shared edges. This model may not give us a distribution of results as output (as in the Monte-Carlo approach) but provides us with the first moment of a distribution with zero confidence interval in a single model run.

There is a series of papers on the topic of mean-field approximation of networks with SIS and SIR dynamics. Pastor-Satorras and Vespignani (2001) proposed a replacement in equations (1) $\rho(t) \rightarrow k\gamma\Theta(\gamma)$ where $\gamma$ is the rate of infection via contact with a single infective individual and $\Theta(\gamma)$ is the probability that the neighbor at the other end of an edge will in fact be infective. Then, replacing $s$ and $\rho(t)$ by $s_k$ and $\rho_k$ (degree-dependent generalizations representing the fraction of vertices of degree $k$ that are susceptible or infective) they rewrote (1) as single differential equation:

$$\frac{d\rho_k}{dt} = k\gamma\Theta(\gamma)(1-\rho_k) - \rho_k \ . \tag{3}$$

The main assumption of this formulation is the same $\Theta(\gamma)$ for all vertices, when in general it depends on the vertex degree. A more detailed approximation is proposed by Volz (2008) who defines the hazard for node with a degree $k$ becoming infected at time $t$ as

$$\lambda_k(t) = rkp_I(t) \ . \tag{4}$$

where $r$ is the infection rate (at which infectious nodes infect a neighbor), $p_I(t) = \frac{M_{SI}}{M_I}$ the probability that an arc from a susceptible node (S) points to an infected node (I). The main disadvantage of this approach is that the probability depends on the node's degree only and not on the parameters of the distribution of outgoing links. For example, if the node has $k$ edges and $k$ is large then there are many options for the number of links shared with infected nodes from 0 to $k$ and thus the probability to be infected will be completely different while the total number of connections is the same. In this case simple averaging is not always possible as the dependence between the number of connections with infected nodes and the probability to be infected is not linear. The main objective of this paper is to modify the well known model (1) by introduction of real-world dynamical network properties into a new system of differential equations and then apply them to obtain a better understanding of the spreading of HIV infections.

## 2 General equations

### 2.1. Network properties

One of the main network properties that have a strong influence on the disease propagation dynamics (and related spreading processes) is the nodes degree distribution. It is well known that many real-world networks have a power-law degree distribution (scale-free networks) (Newman 2003).

$$P(k) \sim k^{-\gamma}, \quad k \leq k_{max}, \tag{5}$$

The exponent $\gamma$ (parameter of the distribution) determines the specific structure of the network. It is important that we restrict $k$ to $k_{max}$ as is the case in real networks i.e. the network has final size. The value $k_{max}$ (highest degree in the network) depends on network size or internal properties of the described system. For example, many studies show that sexual contact networks have scale-free properties (Liljeros 2001) although it is clear that there is a physical threshold for the number of contacts per node in such a network. This fact should not be neglected in the development of a model because it has a strong influence on the simulation results. So our goal is to develop an analytical description of the spreading process over the network using these stochastic considerations. Networks in the examples discussed below have scale-free properties, even though the proposed models may be used for dynamical networks with arbitrary distribution of the degree of nodes.

### 2.2 Link and Node Dynamics

The main idea is to approximate the distribution of the number of links shared between nodes of different types and apply transmission rules independently, not only for every collection of degree-dependent nodes but also for subsets of nodes with the same degree $k$ but different numbers of links shared between nodes of different types (ranges from 0 to $k$). . In terms of epidemiology a node may be connected to infected or to healthy nodes. For instance, if a node has only one link then it can have a shared edge with a healthy or infected node (2 combinations are possible). If there are two links per node then it may have shared edges with healthy or infected nodes or one shared edge with an infected and one with a healthy node (3 combinations), etcetera. Due to a large number of nodes in real complex networks it is possible to generalize this in the form of a binomial (or multinomial) distribution $L(N,k,p) = C_N^k p^k (1-p)^{N-k}$, where $N$ represents the number of links in the collection of nodes with the same degree (ranges from 1 to $k_{max}$), $k$ (ranges from 0 to N) stands for the number of links going out to nodes with specific properties (for example, from healthy to infected) and $p$ indicates the probability to have a specific connection. For heavy tail (large $N$) distributions it makes sense, for computational efficiency, to use the 'de Moivre-Laplace' theorem instead of a binomial distribution. As a rule $p$ is proportional to the fraction of links going out to the group of nodes with specific properties (for example, infected), with $p = \frac{<k_{inf}>}{<k>}$. As $<k_{inf}>$ changes over time the value $p$ is time dependent. Here the node's degree of infected group is neglected as the transmission rate is assumed independent of infected node's degree yet it may be taken into account too. Summing all the combinations of connections multiplied by the probability of a disease transmission we get the number (or fraction) of newly infected nodes for a collection of nodes with the same degree. Summing all the newly infected nodes for all degree-dependent collections of nodes one may find the total number of newly infected nodes. This way it is possible to write down for any collection of nodes with degree $k$ its own set of equations of dynamics:

$$\frac{ds_k(t)}{dt} = -s_k(t) \sum_{l=1}^{k} f(l,\lambda) \cdot L(k,l,p)$$
$$\frac{d\rho_k(t)}{dt} = -\mu\rho_k(t) + s_k(t) \sum_{l=1}^{k} f(l,\lambda) \cdot L(k,l,p), \quad (6)$$
$$\frac{dr(t)}{dt} = \sum_{k} \mu\rho_k(t)$$

where the coefficients $\lambda$ and $\mu$ are infection and removal rates (probabilities) and have the same meaning as in model Eq. 2. The function $f(l,\lambda)$ denotes the removal rates with respect to the number of links $l$ shared with infected nodes. It may be chosen arbitrary and normally has the form $f(l,\lambda) = 1-(1-\lambda)^l$. The value of $p = \frac{<k_{inf}>}{<k>}$ contains information about the node degree distribution and is constant at each time step for all $k$. In equation (4) the network is considered to be infinite. As the model is based on relative values there is no need for a specific number of nodes in the equations and the simulation output can be adjusted to any network size. Function L represents the fraction of nodes having connections to infected nodes with a given degree.

In equation (4) a full rewiring of the network is assumed. This means that we have a new network at each time step $dt$, but with the same node-states. It is assumed that the contact between nodes is random but proportional to degree of the nodes. It is possible to reformulate equation (4) for the network without links dynamics. That leads to a closed system of equations where the fraction of nodes shared between *susceptible* and *infective* nodes will be calculated only for newly infected nodes. The assumption about the links dynamics makes the equations somewhat easier to manipulate and interpret.

A comparison of an agent-based simulation of the reduced model from section 1.2 with steps 1, 2, 6 and a numerical solution of the set of equations (4) is presented in Appendix 1.

## 2.2 Assumptions and limitations

The following assumptions hold for Eq. 6: (i) the population is of constant size; (ii) model parameters $\lambda$ and $\mu$ are constant; (iii) the probability to find a connection with an infected node is proportional to its own degree; (iv) the probability to become infected is proportional to the number of shared edges with an infected node; (v) all the nodes in the network have the same type; (vi) special network properties like community structures or spatial properties are ignored.

The first assumption (population size) is not strict, it can easily be extended taking into account the demography. For instance, it may have the form of $d(s_k(0) - s_k(t))$, where $d$ is a demographic variable. This way we can take into account the number of nodes removed from the network due to natural causes (such as aging) as well as adding nodes to network.

Model parameters $\lambda$ and $\mu$ may vary over time for all of the population or for given groups.

The third assumption refers to the way the links reconfigure over time. The idea of a full reconfiguration of the network at each time step is questionable in many real applications, in this model it can be modified by preserving some of the links in the network. Hence $s_k(t)\sum_{l=1}^{k} f(l,\lambda) \cdot L(k,l,p)$ may be divided into two parts for preserved and renewed links. This approach resembles the configuration algorithm (Bender et al. 1990; Molloy et al. 1995). Our approach allows for adapting the form of the distribution $L(k,l,p)$ to represent any given algorithm of network generation.

The way of transferring infection (or other states changes) can also change according to the field of application; the proposed function is an example which is popular in epidemiology.

**2.3 Consideration of groups with different transmission rate**

By far a more interesting question is how to take into account different types of nodes (different groups), for instance infected nodes in different stages of the disease progression or different genders and its associated transmissibility. For example it is known that for HIV the probability to be infected from man to woman is approximately twice as high as the other way around. Because of this heterogeneity of transmission it is much better to use multinomial distributions. The function $L(k,l,p)$ for one type of nodes then changes to

$$L(k, k_1, k_2, p_1, p_2) = \frac{k_1}{k_1! k_2! k_3} p_1^{k_1} p_2^{k_2} p_3^{k_3}, \qquad (7)$$

where $k_1 + k_2 \leq k$, $k_3 = k - k_1 - k_2$, $p_3 = 1 - p_1 - p_2$, $k$ is number of links of the node, $k_1, k_2$ are the numbers of shared edges with other nodes of type 1 and 2 respectively and $p_1 = \frac{<k>_{type1}}{<k>}, p_2 = \frac{<k>_{type2}}{<k>}$ are the probability to have a connection with an infected node of type 1 and 2. In this case the main equations become a bit more intricate:

$$\frac{ds_k(t)}{dt} = -s_k(t) \sum_{k1=1}^{k} \sum_{k2=1}^{k} f(k_1, k_2, \lambda_1, \lambda_2) \cdot L(k, k_1, k_2, p_1, p_2)_{k_1+k_2 \leq k}$$

$$\frac{d\rho_k(t)}{dt} = -\mu\rho_k(t) + s_k(t) \sum_{k1=1}^{k} \sum_{k2=1}^{k} f(k_1, k_2, \lambda_1, \lambda_2) \cdot L(k, k_1, k_2, p_1, p_2)_{k_1+k_2 \leq k} \quad . \tag{8}$$

$$\frac{dr(t)}{dt} = \sum_k \mu\rho_k(t)$$

The infectious function may be written down as follows: $f(k_1, k_2, \lambda_1, \lambda_1) = 1 - (1-\lambda_1)^{k_1}(1-\lambda_2)^{k_2}$, where $\lambda_1, \lambda_2$ are the transmission probability for group 1 and group 2. This approach may be extended to any number of groups, but for distributions with a very heavy tail it may be computationally inefficient. Nevertheless computations may be more efficient by predefining the coefficient of the multinomial distribution. These coefficients can be stored once, independently of the model. In addition to this approach it is possible to modify the equations for bipartite graphs. In this case all the equations should be divided into pairs with a corresponding change of $p$:

$$\frac{ds_k^1(t)}{dt} = -s_k^1(t) \sum_{l=1}^{k} f(l, \lambda_{2 \to 1}) \cdot L(k, l, \frac{<k_{inf}^2>}{<k>})$$

$$\frac{ds_k^2(t)}{dt} = -s_k^2(t) \sum_{l=1}^{k} f(l, \lambda_{1 \to 2}) \cdot L(k, l, \frac{<k_{inf}^1>}{<k>})$$

$$\frac{d\rho_k^1(t)}{dt} = -\mu\rho_k^1(t) + s_k^1(t) \sum_{l=1}^{k} f(l, \lambda_{2 \to 1}) \cdot L(k, l, \frac{<k_{inf}^2>}{<k>}) \quad , \tag{9}$$

$$\frac{d\rho_k^2(t)}{dt} = -\mu\rho_k^2(t) + s_k^2(t) \sum_{l=1}^{k} f(l, \lambda_{1 \to 2}) \cdot L(k, l, \frac{<k_{inf}^1>}{<k>})$$

$$\frac{dr(t)}{dt} = \sum_k \mu\rho_k^1(t) + \sum_k \mu\rho_k^2(t)$$

where the superscript index at $s$ and $\rho$ denotes the different parts of the bipartite graph and $\lambda_{1 \to 2}, \lambda_{2 \to 1}$ are the transmission probabilities from the first type of nodes to the second and vice versa.

## 2.4 Phase diagrams of the stochastic network dynamics

Phase plots are an intuitive way to analyze dynamical systems. In Figure 1 we show nine phase plots representing the dynamics of infected nodes with different degree.

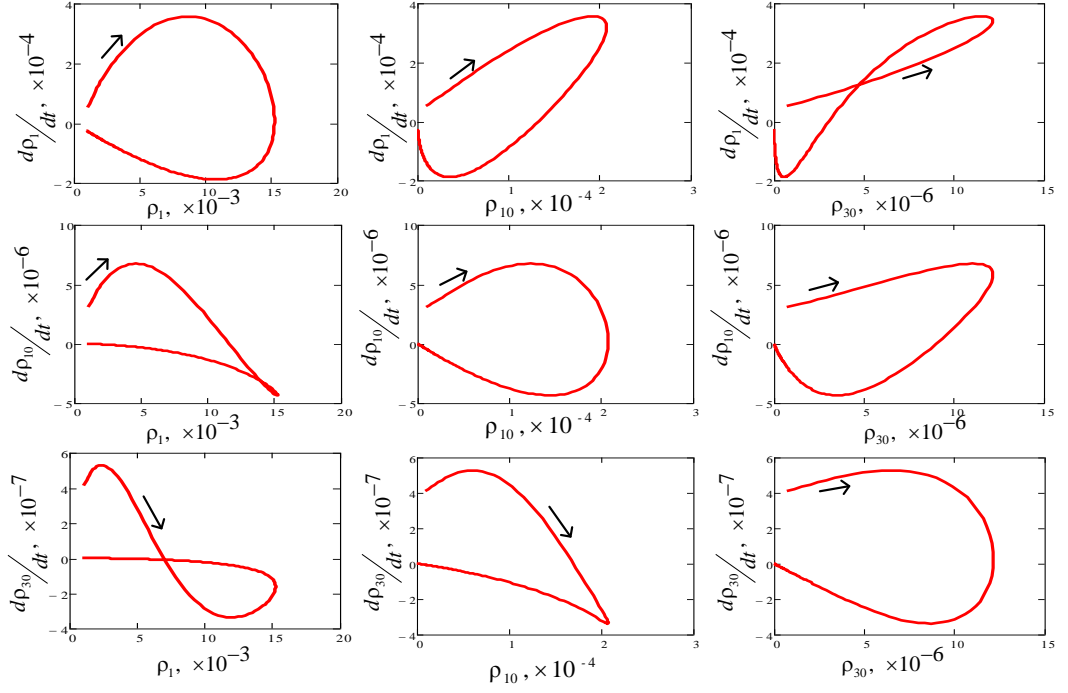

**Fig. 1**. Phase plots for infected nodes using Eq. 4. Parameters used are: $\lambda = \mu = 0.05$, $\rho(0) = 0.01$, with power-law degree distribution $\gamma = 3$, $k_{max} = 60$.

Fig. 1 indicates how the power-law degree changes the shape of phase trajectory of nodes with different degree $\left(\rho_m, \dfrac{\rho_n}{dt}\right)$, where m=n, but due to a different rate of infection spreading for nodes with different degree the phase plot is twisted for $\left(\rho_m, \dfrac{d\rho_n}{dt}\right)$, where m $\neq$ n. In this case the sign of the derivative may change several times.

Here if we would not interpret group $r$ as removed but rather as recovered with immunization (for model it is the same because group $r$ doesn't facilitate the infection spreading), a new dynamics for the healthy (not infected) group ($s+r$) can be studied (see Fig.2).

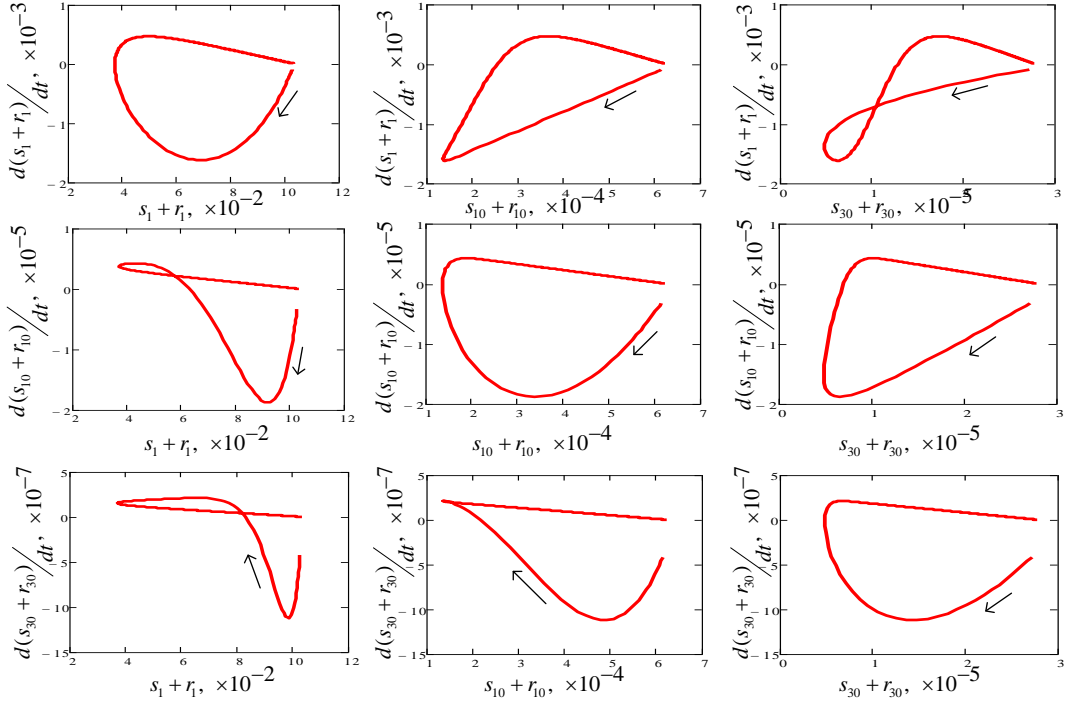

**Fig. 2**. Phase plots for healthy nodes in Eq. 4. Parameters used are $\lambda = \mu = 0.05$, $\rho(0) = 0.01$, with power-law degree distribution, $\gamma = 3$, $k_{max} = 60$

There is an interesting effect at the later stages of the process when the epidemic is close to dying out and there emerges a linear dependency between the size of the healthy group and its derivative. Curves are only shown for 3 sets of nodes ($k = 1$, 10 and 30) as the different behavior for them can be expected. It is clear that the diagonal phase plot in Figures 1 and 2 have the same shape but a different size which decreases with increasing $k$. At the same the cross-influence of nodes with different degree is shown to be very nonlinear.

## 2.5. Sensitivity analysis

Using sensitivity analyses we study the model response to changes in selected model parameters (Saltelli et al. 1999; Sobol 1990). As our model has no analytical solution, we use a sampling based method that involves running the original model for a set of input parameters and estimating the sensitivity of the model outputs at those sample points. Let's consider sensitivity indices for every output point in the form

$$S_j = \frac{Var[E(Y | x_j)]}{Var(Y)}, \quad (10)$$

where $S_j$ is the global sensitivity index for variable $x_j$ and $Y$ is an output of the model. The most interesting parameters for analysis are parameters of the distribution (such as the power-law distribution exponent), infection probability and initial fraction of infected nodes. The drifting of indices for those parameters for every time point of model output is shown in Fig. 3.

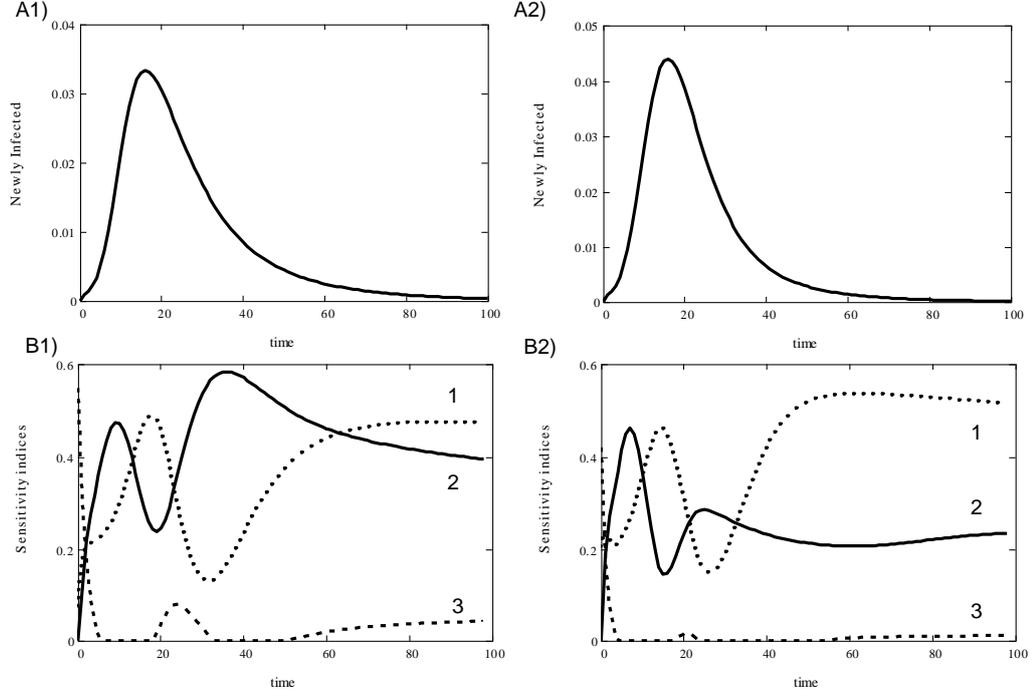

**Fig 3.** Drifting of sensitivity indices for the power-law degree distribution exponent $\gamma$ (dotted line 1), infection probability $\lambda$ (solid line 2) and initial fraction of infected nodes $\rho_0$ (dashed line 3); A1) $\gamma = 2.5$, $\lambda = 0.1$, $\rho(0) = 0.005$, A2) $\gamma \in U(2,3)$, $\lambda \in U(0.05, 0.15)$, $\rho(0) \in U(0.001, 0.01)$; B1) $\gamma = 2.75$, $\lambda = 0.15$, $\rho(0) = 0.005$, B2) $\gamma \in U(2, 3.5)$, $\lambda \in U(0.05, 0.25)$, $\rho(0) \in U(0.001, 0.01)$

Fig. 3 clearly shows that the influence of the network structure on the final variance of the model output is substantial and normally changes in the range of 20% to 50%. The largest impact is observed in stages of fast growth and drops of for newly infected nodes. The influence of the initial fraction of infected nodes is considerable only in the first steps of the simulation. The peak of incidence is mostly controlled by the infection probability.

# 3 Understanding the historical HIV-AIDS outbreak dynamics

The model presented in Eqs. 6 and 7 has been adapted to study the case of the HIV/AIDS epidemics outbreak through sexual contact networks (Sloot et al.

2008; Mei et al. 2011). Two exposure groups were studied: a homosexual (Men having Sex with Men; MSM) and a heterosexual population. Parameters used, $\gamma$, $k_{max}$, $\lambda$, $I_0$ for both populations as well as for the historical data are taken from the USA population and were provided by the Virolab consortium (Virolab). For identification of model parameters we used the official estimation of HIV/AIDS cases as reported by the Center for Disease Control CDC[1]. Two different types of infected nodes are modeled for both the Heterosexual and the MSM populations. One type represents an infected node without treatment and a second type represents a node with treatment (AZT or even Highly Active Antiretroviral Therapies: HAART). Individual disease progression is calculated using a Markov chain model with different infection probabilities for each stage of infection in each subpopulation. The Markov chain model (Sweeting et. al 2005) was modified to be used in a set of ODEs in the form of transmission rates without any stochasticity. This means that we calculate the fraction of people in a new state at each time-step of the numerical solution. The final state of the Markov chain model (death) is equivalent to the removal rates $\mu$ in the general SIR model (see Eq. 4). The model is not stationary due to changes over time induced by new diagnosis and treatment, notably the introduction of HAART in 1996. For the two populations the model has a form given by Eq. 7 where the two parts of the network are men and women respectively. Simulation results together with the historical data (Virolab) are shown in the phase plots of Fig 4. Some implementation details of the model in the form of an ODE are presented in Appendix 2.

---

[1] http://www.cdc.gov/hiv/

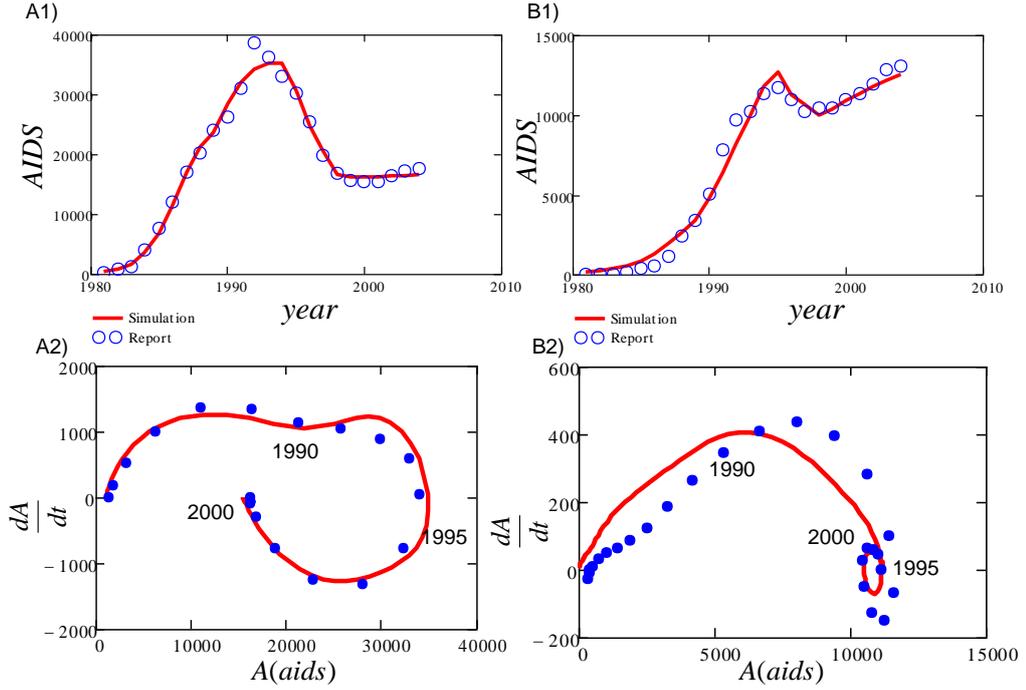

**Fig. 4** Simulation results (drawn lines) and historical data (circles and dots) for the AIDS epidemic in USA. A1) Incidence for MSM, A2) Phase plot for MSM. B1) Incidence for heterosexual population B3) Phase plot for the heterosexual population.

Thanks to the deterministic character of ODEs introduced in this paper and the related smooth derivatives in the simulation results, we now can make use of efficient optimization algorithms like fast gradient descent methods for fine tuning and identification of the model parameters. Also we can study in more detail the phase plots per set of nodes, each with a different degree. Some results are shown in Figs. 5 and 6.

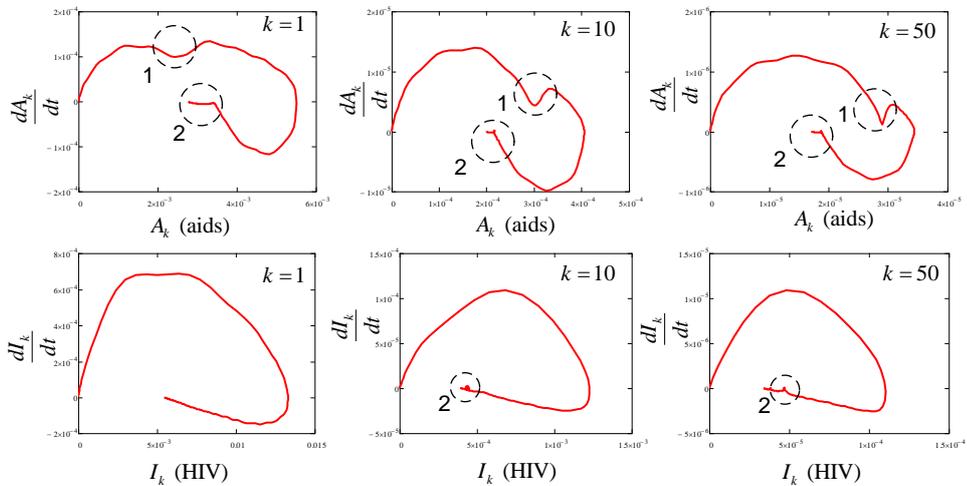

**Fig. 5** Simulation results for the AIDS epidemic in USA (MSM group), using a power-law distribution with $\gamma = 1.6$ and $k_{max} = 250$, $I_0 = 0.32\%$, $\lambda = 0.44$ phase plots

From Fig. 5 we see an interesting effect happening at the start of treatment (circle 1, 1988) and at the introduction of HAART (circle 2, 1996). The influence of these factors on the nodes dynamics with different degree is not the same. The changes in model coefficients lead to more pronounced changes in the dynamics for nodes with higher degree. This is a new finding that has not been reported before to the best of our knowledge. The flat zone at the late stages of the epidemics is a stabilization of incidence mostly due to demographic factors which prevent the disappearance of the epidemic. The same effect can be observed from Fig. 6 where the phase plots for the heterosexual population are shown.

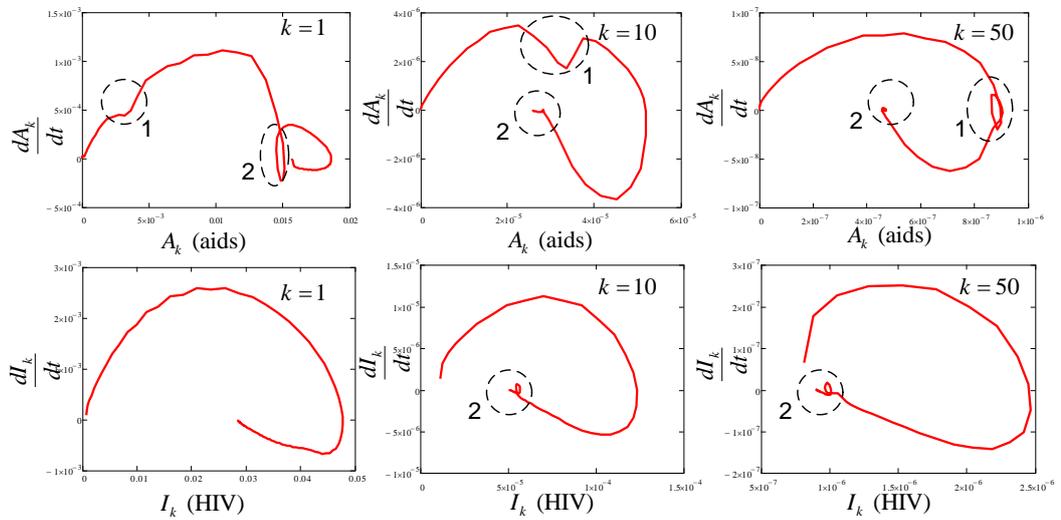

**Fig. 6.** Simulation results for the AIDS epidemic in USA (heterosexual group), power-law distribution with $\gamma = 2.7$ and $k_{max} = 60$, $I_0 = 0.2\%$, $\lambda = 0.28$, phase plots.

Note that the phase plots for heterosexual population differ from those for MSM. The main difference is due to a difference in the response to treatment. The knots in Fig. 6 ($k=1$, $k=50$) are a consequence of the lack of convergence in the derivation of a natural run of the epidemics and the additional effect due to treatment. In case of coincidence one can see a temporal drop of the derivation value (see plot for MSM AIDS, Fig. 5). This difference for MSM and heterosexual population dynamics can be explained by the different network properties for those populations.

# Conclusions

In this paper we introduced a system of nonlinear ordinary differential equations with stochastic terms to model the dynamics of spreading processes, with an emphasis on the edge dynamics, in random heterogeneous networks. The models were verified numerically and show good computational efficiency. For direct simulation and relatively large networks however (~1 million of nodes), computations may take from tens of minutes for one statistically significant experiment up to several days for specific calculations like optimization or sensitivity analysis. For the introduced ODE approximation the computational time doesn't depend on network size but is only sensitive to tail heaviness of nodes degree distribution and the number of groups with different transmission probabilities. If those values are not extremely large the computational time should not exceed a few seconds. Moreover it reduces the time to build new network models with different parameters and properties. This new approach was used to study different cases like bipartite graphs or graphs with several disparate types of nodes. This approach allowed us to study the dynamics using the regular type of analysis for systems of differential equations (phase plots, sensitivity analysis) were made as well. A new form of analysis of the epidemiological network models is proposed by the example of HIV. As one consequence of the expressiveness of the model introduced, we observe new dynamics in the HIV outbreak and identify pronounced changes in the dynamics for nodes with higher degree in the MSM populations at points in time where the treatment changes. Our approach allows for tuning of the parameters of the historical data model using efficient optimization algorithms. The principles of deterministic model building are quite general and may be extended to networks with any node degree distribution or underlying statistical properties.

# Acknowledgement

The authors would like to acknowledge the financial support of the European Union through the ViroLab project (www.virolab.org), EU project number: IST-027446 and the Dynanets Project (www.dynanets.org), EU Grant Agreement Number 233847. The research was partly sponsored by a grant from the Leading Scientist Program of the Government of the Russian Federation, under contract 11.G34.31.0019 and "Multi-Disciplinary Technological Platform for Distributed Cloud Computing Environment Building and Management CLAVIRE" performed under Decree 218 of Government of the Russian Federation. PS also acknowledges support from the NTU complexity program.

# Appendix 1

Comparison of models: result of 100 agent-based simulations and the results of the numerical solution of the set of ordinary differential equations.

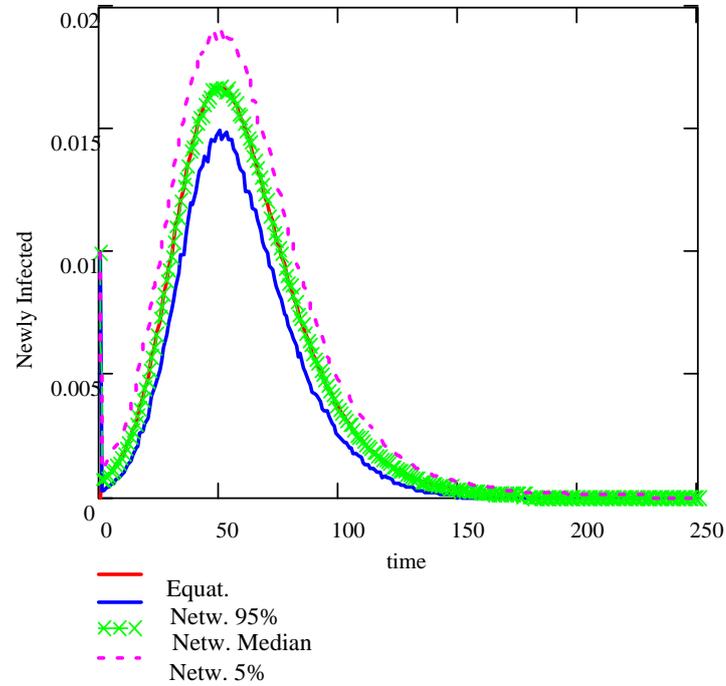

**Fig. 1**. Spreading process on random network: $P(k) \sim k^{-\gamma}$, $\gamma = 3.0$, $K_{max} = 30$, $i = 0.05\,(infection\ rate)$, $Network\ Size = 10^4$

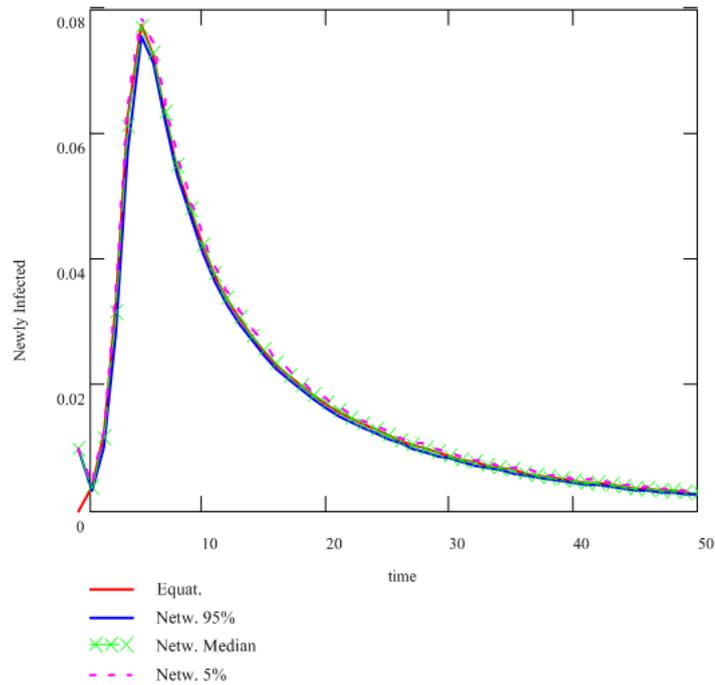

**Fig. 2.** Spreading process on random network: $P(k) \sim k^{-\gamma}$, $\gamma = 1.6$, $K_{max} = 150$, $i = 0.05\,(infection\ rate)$, $Network\ Size = 10^5$

# Appendix 2

This section presents the equations of HIV dynamics. Some details of HIV progression model are skipped and may be found in (Sloot et al. 2008) of main reference list.

The equations for homosexual populations are

$$\frac{ds_k(t)}{dt} = -s_k(t)\sum_{k1=1}^{k}\sum_{k2=1}^{k} f(k1,k2,i) \cdot L(k,k1,k2,p1,p2)_{k1+k2 \leq k} + ds_k(0) - ds_k(t)$$

$$\frac{dI_k(t)}{dt} = -dI_k(t) + s_k(t)\sum_{k1=1}^{k}\sum_{k2=1}^{k} f(k1,k2,i) \cdot L(k,k1,k2,p1,p2)_{k1+k2 \leq k}$$

$$\frac{dIr(t)}{dt} = \sum_k dI_k(t)$$

The equations for heterosexual populations are:

$$\frac{dsm_k(t)}{dt} = -sm_k(t)\sum_{k1=1}^{k}\sum_{k2=1}^{k} f(k1,k2,0.5i) \cdot L(k,k1,k2,p1_w,p2_w)_{k1+k2 \leq k} + dsm_k(0) - dsm_k(t)$$

$$\frac{dsw_k(t)}{dt} = -sw_k(t)\sum_{k1=1}^{k}\sum_{k2=1}^{k} f(k1,k2,i) \cdot L(k,k1,k2,p1_m,p2_m)_{k1+k2 \leq k} + dsw_k(0) - dsw_k(t)$$

$$\frac{dIm_k(t)}{dt} = -dIm_k(t) + sm_k(t)\sum_{k1=1}^{k}\sum_{k2=1}^{k} f(k1,k2,0.5i) \cdot L(k,k1,k2,p1_w,p2_w)_{k1+k2 \leq k}$$

$$\frac{dIw_k(t)}{dt} = -dIw_k(t) + sw_k(t)\sum_{k1=1}^{k}\sum_{k2=1}^{k} f(k1,k2,i) \cdot L(k,k1,k2,p1_m,p2_m)_{k1+k2 \leq k}$$

$$\frac{dIr(t)}{dt} = \sum_k dIw_k(t) + \sum_k dIm_k(t)$$

Notation and parameters of the models:

- $S_k, I_k$ - fraction of susceptible and infected nodes having k edges (degree distribution)
- $Ir$ fraction of nodes with HIV, removed from the network due to demographic factor
- $d$ demographic coefficient
- $p1 = \frac{<k_{\inf}>_{notreatment}}{<k>}$, probability to have a connection with infected node without treatment
- $p2 = \frac{<k_{\inf}>_{treatment}}{<k>}$, probability to have a connection with infected node with treatment
- $f(k1,k2,i) = 1-(1-i)^{k1}(1-0.4i)^{k2}$, probability to be infected with rate $i$ being edged by $k1$ infected nodes without treatment and $k2$ infected nodes with treatment. $0.4i$ denotes that probability to be infected from infected nodes with treatment is 40% less.
- $L(k,k1,k2,p1,p2) = \begin{vmatrix} p3 = 1-p1-p2 \\ k3 = k-k1-k2 \\ \text{return } \frac{k!}{k1!k2!k3!} p1^{k1} p2^{k2} p3^{k3} \end{vmatrix}$ , probability to have $k1$ shared edges with nodes without treatment and $k2$ infected nodes with treatment.
- $K_{\min} \leq k \leq K_{\max}$
- Subscripts $m$ and $w$ denote the men and woman population in the heterosexual HIV model.
- The probability to be infected from man to woman is considered twice as high as the other way around: $f(k1,k2,0.5i)$